\documentclass[10pt]{article}
\usepackage{float}
\usepackage[symbol]{footmisc} 
\usepackage[superscript,sort]{cite}
\usepackage{array}
\usepackage{bm}

\newcolumntype{x}[1]{%
>{\centering\hspace{0pt}}p{#1}}%
\usepackage[normalem]{ulem} 
\usepackage[colorlinks=true,linkcolor=black,citecolor=black,urlcolor=black,bookmarks=true,breaklinks=true]{hyperref}
\usepackage[usenames]{color}
\usepackage{amsmath,amssymb}

\usepackage{amsthm,amscd,amsxtra,amsfonts,amsmath,amssymb,multirow}
\usepackage{wrapfig}
\usepackage[footnotesize]{caption}
\usepackage{subcaption}
\usepackage[tiny,compact]{titlesec}
\usepackage{threeparttable} 
\usepackage{helvet}

\usepackage{booktabs}
\usepackage[table]{xcolor}
\usepackage{xcolor,colortbl}
\usepackage{tikz}
\usetikzlibrary{shapes,arrows}
\usepackage[T1]{fontenc}
\usepackage[linesnumbered,ruled]{algorithm2e} 
\titlespacing*{\section}{0pt}{*0}{*0}
\titlespacing*{\subsection}{0pt}{*0}{*0}
\titlespacing*{\subsubsection}{0pt}{*0}{*0} 
\titlespacing{\paragraph}{0pt}{*0}{*1}

\definecolor{MyPurple}{rgb}{1,0,1}

\newcommand{\beq}[1]{\begin{equation} \label{#1}}
\newcommand{\eeq}{\end{equation}}
\newcommand{\barray}{\begin{array}{ll}}
\newcommand{\earray}{\end{array}}

\begin{document}
\pagenumbering{roman}

\clearpage \pagebreak \setcounter{page}{1}
\renewcommand{\thepage}{{\arabic{page}}}

\title{Topological fingerprints reveal protein-ligand binding mechanism
}

\author{
Zixuan Cang$^1$,
 and
Guo-Wei Wei$^{1,2,3}$ \footnote{ Address correspondences  to Guo-Wei Wei. E-mail:wei@math.msu.edu}\\
$^1$ Department of Mathematics \\
Michigan State University, MI 48824, USA\\
$^2$  Department of Biochemistry and Molecular Biology\\
Michigan State University, MI 48824, USA \\
$^3$ Department of Electrical and Computer Engineering \\
Michigan State University, MI 48824, USA \\
}

\date{}
\maketitle
\maketitle
\abstract{
Protein-ligand binding is a fundamental biological process that is paramount to many other biological processes, such as signal transduction, metabolic pathways, enzyme  construction, cell secretion, gene expression, etc. 
Accurate prediction of protein-ligand binding affinities is vital to rational drug design and the understanding of protein-ligand binding and binding induced function. 
Existing binding affinity prediction methods are  inundated with  geometric detail and involve excessively high dimensions, which undermines their predictive power for massive binding data. Topology provides an ultimate level of abstraction and thus incurs
 too much reduction in geometric information. 
Persistent homology   
embeds geometric information into topological invariants and bridges the gap between complex geometry and abstract topology.  However, it  over simplifies biological information.  
This work introduces element specific persistent homology (ESPH) to retain crucial biological information during topological simplification. 
The combination of ESPH and machine learning gives rise to one of the most efficient and powerful tools for revealing protein-ligand binding mechanism
and for predicting  binding affinities. 
}

\section{Introduction}
The study of protein-ligand binding has attracted enormous attention in the past few decades due to its importance to biochemistry, biophysics and biomedicine \cite{Kollman:2000ACR,gilson1997statistical, Gilson:2007}. Experimentally, three-dimensional (3D) structures of protein-ligand complexes  obtained from X-ray crystallography, cryo-electron microscopy, and nuclear magnetic resonance  shed  light on molecular recognition, protein conformational changes upon binding, and possible allosteric effect.  Sophisticated physical and chemical tools, including quantum mechanical calculation, molecular dynamics simulation, and Monte Carlo sampling have been applied to protein-ligand binding analysis.  Nevertheless, despite of much effort, our ability to predict binding affinities is still quite limited, which suggests a significant gap in our understanding. 
  
Essentially, there are three types of scoring functions for protein-ligand binding predictions: physics based \cite{Ortiz:1995,Yin:2008}, knowledge based \cite{HongjianLi:2014RF} and empirical ones \cite{Zheng:2015LISA,Verkhivker:1995PLP, Eldridge:1997,WangRenXiao:2002}.  In general, physics based scoring functions invoke QM and QM/MM approaches to provide unique insights into the molecular mechanism of protein-ligand interactions. 
A prevalent view is that binding involves intermolecular forces, such as steric contacts, ionic bonds, hydrogen bonds, hydrophobic effects and van der Waals interactions. Meanwhile, empirical scoring functions might work well with carefully selected data sets. However, both  physics based scoring functions and empirical scoring functions are not designed to deal with increasingly diverse and rapidly growing data sets.
Additionally,  most data sets of protein-ligand complexes generated from x-ray crystallography have limited resolutions around two \AA~ and typically do not include hydrogen atoms. NMR structures have less resolutions in general. For this type of data sets, physics based models often contain too much geometric detail that are  not available from experimental data. Knowledge based scoring functions resort to modern machine learning techniques, which utilize nonlinear regressions and exploit large data sets to uncover  hidden patterns in the data sets, and are able to outperform other scoring functions in massive and complex data challenges. However,  the data-driven feature selections in knowledge based scoring functions tend to abandon physical consideration  and render a high dimensional problem.  
 
In this work, we propose an entirely new strategy that integrates  persistent homology \cite{Edelsbrunner:2002,Zomorodian:2005} 
and machine learning to elucidate molecular mechanism in protein-ligand binding and  predict  binding affinities. Mathematically, both physics based scoring functions and empirical ones are geometry based models that are often inundated with too much  structural detail resting in excessively high dimensions. In contrast, topology  deals with the connectivity of different components in a space, and  characterizes independent entities,  rings and higher dimensional topological faces within the space \cite{kaczynski:mischaikow:mrozek:04}. It provides the ultimate level of abstraction of many biological processes, such as the  open or close of ion channels, the assembly or disassembly of virus capsids, the folding and unfolding of proteins and the association or disassociation of ligands. However, conventional topology or homology is truly free of metrics or coordinates, and thus retains too little geometric information to be practically useful. Persistent homology   is a new branch of algebraic topology that   embeds multiscale geometric information into topological invariants to achieve the interplay between geometry and topology. However, it over simplifies biological information to be useful in quantitative predictions. We  introduce element specific persistent homology (ESPH) to dramatically reduce biomolecular complexity while retaining crucial chemical and biological information. ESPH enables us to decipher the entangling code of protein-ligand interactions. We also propose interactive persistent homology (IPH) to describe the interactions between protein and ligand. Finally, we develop binned barcode representation (BBR) to characterize the strength of various protein-ligand  interactions. 
The proposed ESPH, IPH and BBR give rise to  an appropriate level of topological abstraction, geometric specificity, and biological information to reveal the molecular mechanism of protein-ligand binding and deliver a concise, efficient, accurate while low dimensional representation of protein-ligand interactions. The barcodes generated by the persistent homology computation are called topological fingerprints (TF) or element specific topological fingerprints (ESTF) if ESPH is used. The features to be fed to machine learning methods are generated from TFs/ESTFs using various treatments including BBR. We demonstrate that our topology based scoring function (T-Score) outperforms all the other existing  methods in binding affinity predictions of two large benchmark data sets.

\section{Methods}
\subsection{Persistent homology analysis of protein-ligand complexes}

\begin{figure}[ht]
\begin{center}
\includegraphics[keepaspectratio,width=0.5\columnwidth]{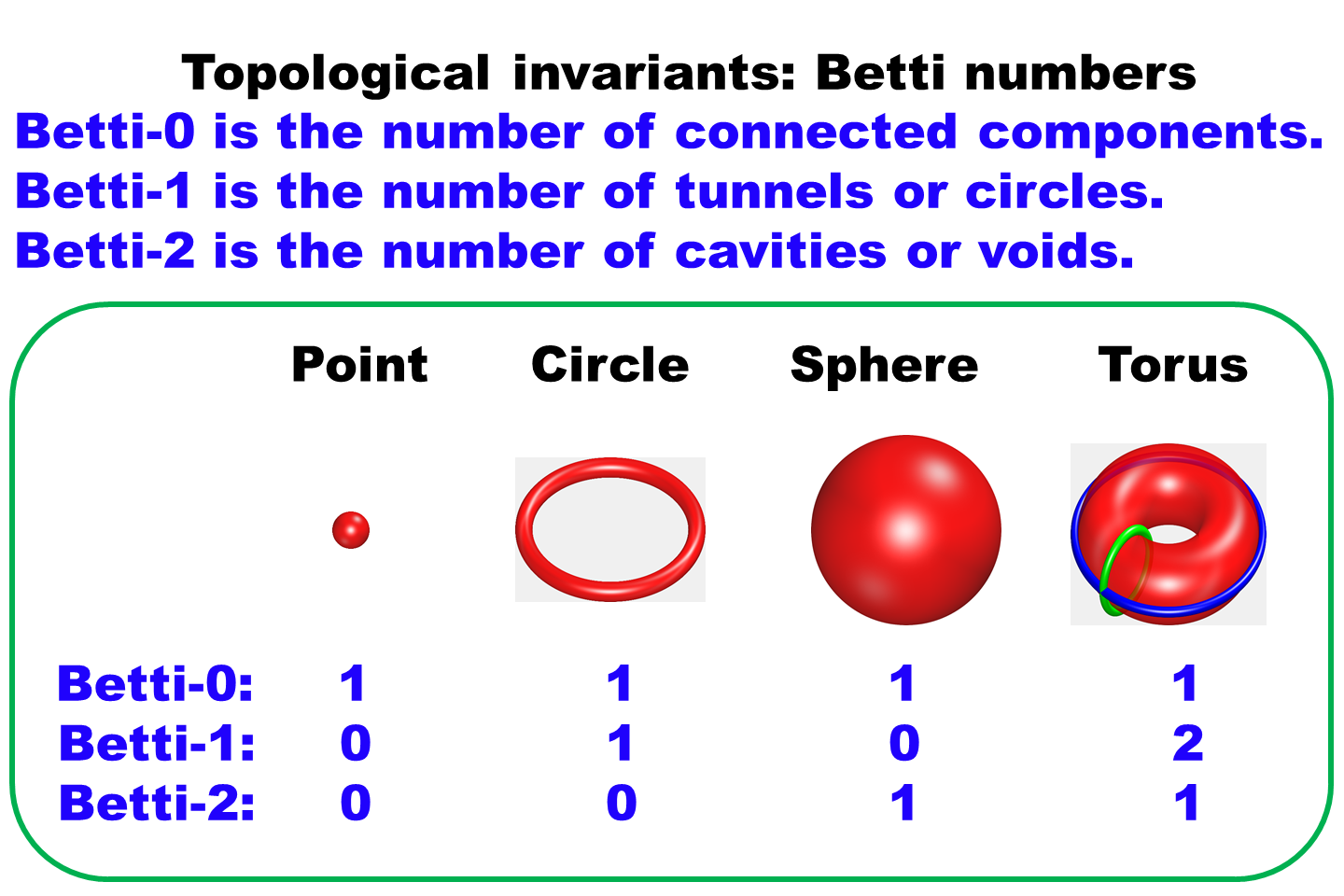}
\caption{An illustration of topological invariants, i.e., Betti numbers  for a point, a circle, an empty sphere and a torus. For the torus, two auxiliary rings are added to explain Betti-1$=2$.        }
\label{fig:invariants}
\end{center}
\end{figure}

\begin{figure}[ht]
\begin{center}
\includegraphics[keepaspectratio,width=0.7\columnwidth]{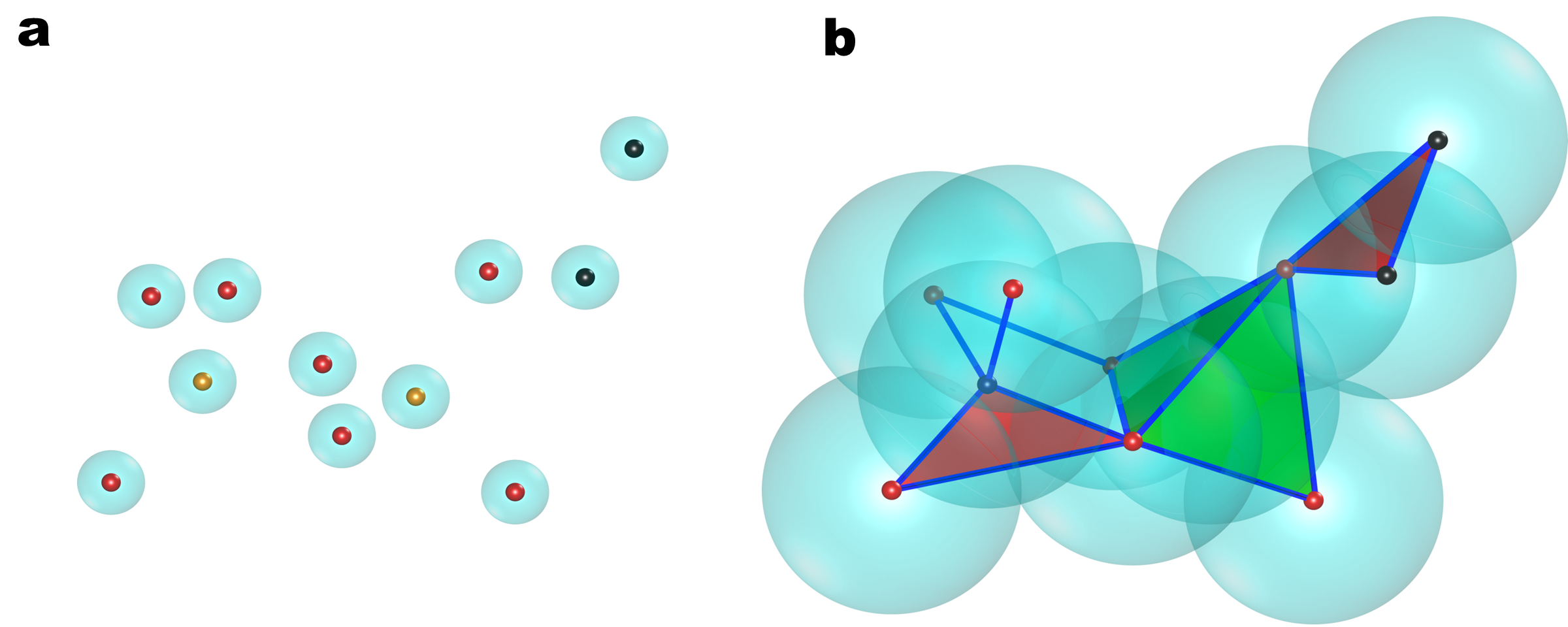}
\caption{An illustration of simplicial complex and filtration of the ligand heavy atoms in protein 3LPL. Red balls are oxygen atoms, black balls are carbon atoms, and orange balls are phosphorus atoms. 
{\bf a}:  Positions of   heavy atom spheres are generated via the radius filtration at $r=0.7$\AA. In this case, one has eleven 0-simplexes, zero 1-simplex and zero 2-simplex.        
{\bf b}: Filtration progresses to $r=2.56$\AA. In this case, one has zero 0-simplex, three 1-simplex, two 2-simplexes and one 3-simplex.  
}
\label{Figure1}
\end{center}
\end{figure}

The fundamental task of topological data analysis is to extract topological invariants, namely the intrinsic features of the underlying space, of a given data set without additional structure information, like covalent bonds, hydrogen bonds, van der Waals interactions, etc.
A fundamental  concept in  algebraic topology is simplicial homology, which concerns the identification of topological invariants from a set of discrete nodes such as atomic coordinates in a  protein-ligand complex. For a given  configuration, independent components, rings and cavities are topological invariants and their numbers  are called Betti-0, Betti-1 and Betti-2, respectively, see Fig. \ref{fig:invariants}. To  study topological invariants in a discrete data set,  simplicial homology uses  a specific rule such as Vietoris-Rips (VR) complex, C$\check{e}$ch complex or alpha complex to identify simplicial complexes.  Specifically, a 0-simplex is a vertex, a 1-simplex an edge, a 2-simplex  a triangle, and a 3-simplex represents a tetrahedron, see Fig. \ref{Figure1}. Algebraic groups built on these simplicial complexes are employed in simplicial homology to systematically compute various Betti numbers.

Simplicial homology  is metric free and thus is too abstract to be insightful for complex and large protein-ligand binding data sets. Persistent homology consists of a series of homologies constructed over a filtration process, in which the connectivity of the given data set is systematically reset according to a scale parameter.  In the  Euclidean-distance based filtration for biomolecular coordinates,  the scale parameter is an ever-increasing  radius of  an ever-growing ball whose center is the coordinate of each atom, see Fig. \ref{Figure1}.  Therefore, filtration induced persistent homology gives a multiscale representation of the corresponding topological space and reveals topological persistence of  the given data set. 
 

\begin{figure}[ht]
\begin{center}
\includegraphics[keepaspectratio,width=0.7\columnwidth]{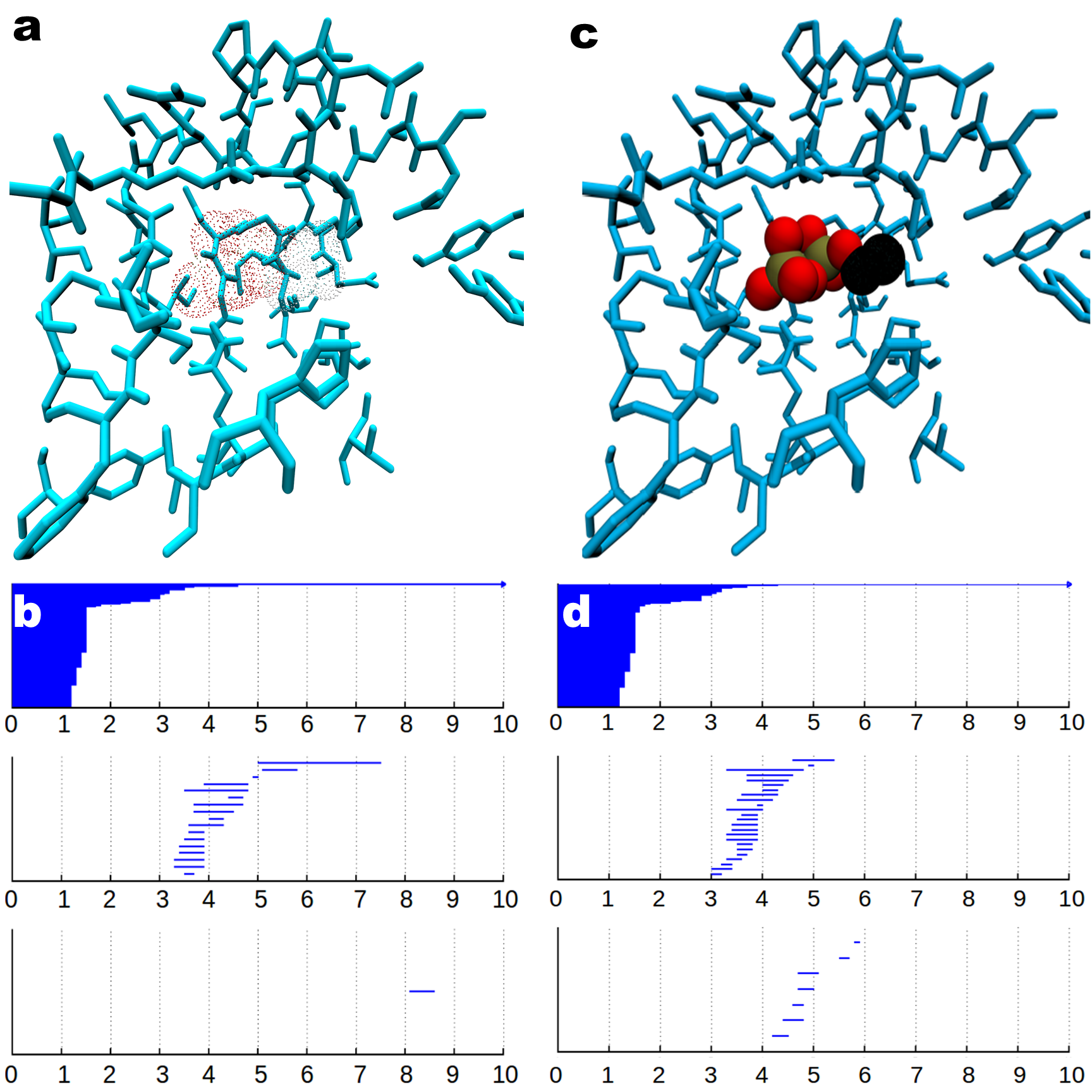}
\caption{Illustration of  protein-ligand binding induced TF change. The hydrogen atoms are shown but not used in TF computation.
{\bf a}: Binding site residues of protein 3LPL.
{\bf b}: The TFs of the heavy atoms of protein 3LPL binding site residues without the ligand.   
{\bf c}: Binding site residues of protein 3LPL and the ligand.
{\bf d}: The TFs of the heavy atoms of protein 3LPL binding site residues and the ligand.
  }
\label{Figure2}
\end{center}
\end{figure}

The power of persistent homology lies in its topological abstraction and dimensionality reduction. It reveals the topological connectivity in 
biomolecular complexes in terms of   topological fingerprints (TFs) \cite{YaoY:2009, KLXia:2014c, Mate:2014, ZXCang:2015}, which are recorded  as  the barcodes \cite{Ghrist:2008} of biomolecular topological invariants over the filtration.  It is worthy to mention that topological connectivity differs from chemical bonds, van der Waals bonds or hydrogen bonds.  Indeed, TFs  offer an entirely new  representation of protein-ligand interactions. 
Figure \ref{Figure2} depicts the TFs of protein-ligand complex 3LPL. By a comparison of TFs of the protein   and those of the corresponding   protein-ligand complex near the binding site,  the changes in Betti-0, Betti-1 and Betti-2 panels can be easily noticed. For example, more bars occur in the Betti-1  panel  around the filtration parameter values 3-5\AA ~ after the binding, which indicates the potential hydrogen bonding network due to the protein-ligand binding. Additionally, the binding induced Betti-2 bars in the range  of 4-6\AA~ reflect potential protein-ligand hydrophobic contacts. In fact, the change in Betti-0 bars is associated with ligand atomic types and atomic numbers. Therefore, TFs and their changes describe the protein-ligand binding in terms of topological invariants.    
  
\begin{figure}[ht]
\begin{center}
\includegraphics[keepaspectratio,width=0.7\columnwidth]{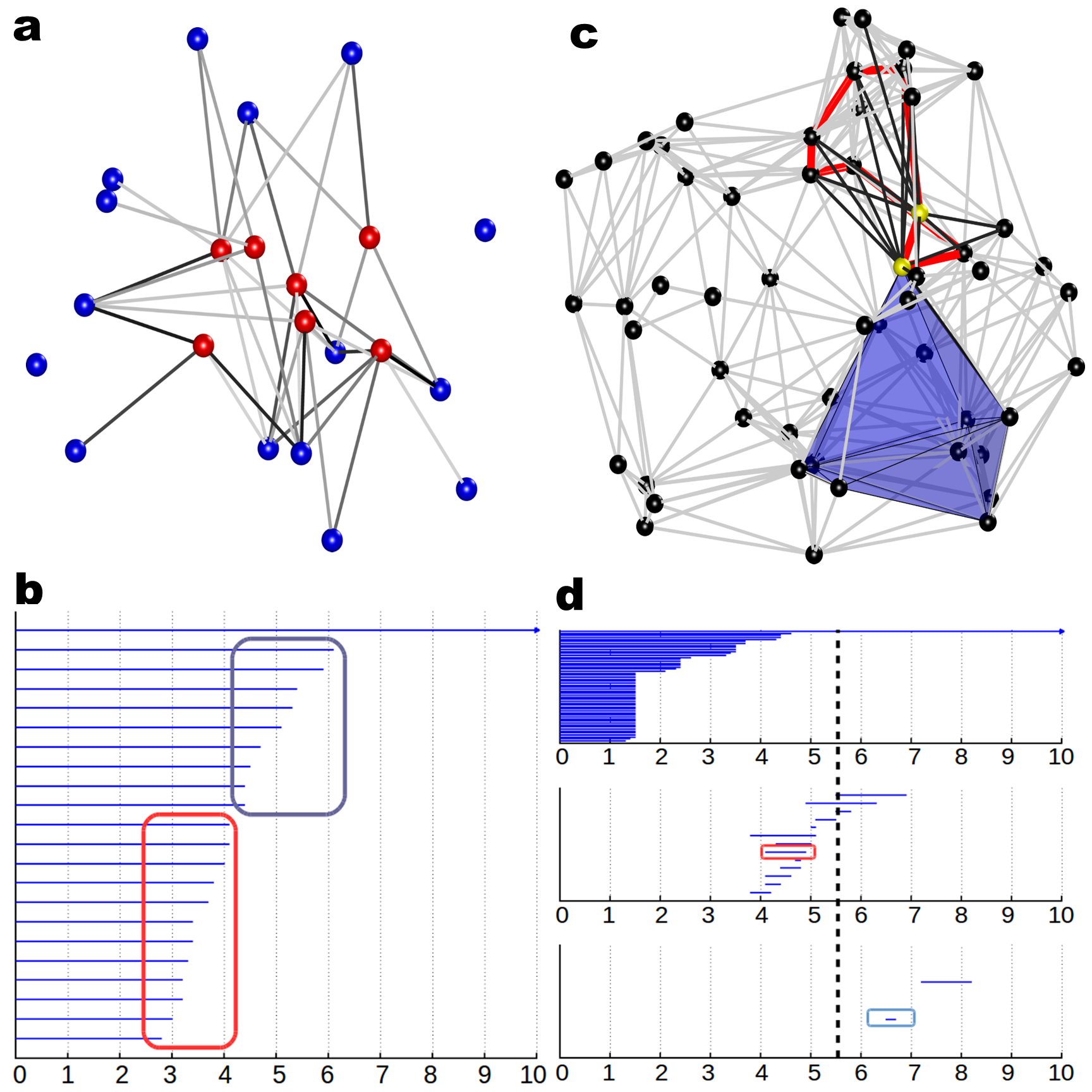}
\caption{An illustration of element specific topological fingerprints (ESTF) indicating the hydrophilic network (Left) and hydrophobic network (Right).   
{\bf a}: Hydrophilic network showing the connectivity  between nitrogen atoms of the protein (blue) and oxygen atoms of the ligand (red). 
{\bf b}:  The Betti-0 ESTFs of the aforementioned ${\rm N-O}$ hydrophilic network.   Betti-0 barcodes show not only the number and strength of 
of  hydrogen bonds, but also the hydrophilic environment. Specifically, bars in  the left   box can be directly  interpreted as moderate or weak hydrogen bonds, while the bars in the right box contributing the degree of hydrophilicity at the binding site. The black dash line shows the corresponding cutoff value of the connection network in {\bf a}.
{\bf c}: Hydrophobic network constructed with carbon atoms near the binding site. 
{\bf d}:  The ESTFs of  {\bf c} highlighting  the protein-ligand hydrophobic contacts. The bars in the red box and the blue box correspond to the red loop and the blue cavity respectively in {\bf c}. The black dash line shows the corresponding cutoff value of the connection network in {\bf c}.
}
\label{Figure3}
\end{center}
\end{figure}

However these collective TFs ignore crucial biological information and  have a limited power in characterizing biomolecules.  To characterize biomolecular systems, we introduce element specific persistent homology (ESPH). Specifically, we consider commonly occurring heavy element types in a protein ligand complex, namely, ${\rm C, N, O,}$ and ${\rm S}$ in proteins and ${\rm C, N, O, S, P, F, Cl, Br,}$ and  ${\rm I}$ in ligands. Our ESPH reduces the biomolecular complexity by disregarding individual atomic character, while retaining vital biological information by distinguishing element types. Additionally, to characterize protein-ligand interactions, we introduce interactive persistent homology (IPH) by  selecting a set of heavy atoms involving a pair of element types, one from protein and the other from ligand, within a given cutoff distance. The resulting  TFs, called interactive element specific topological fingerprints (ESTFs), are able to characterize intricate protein-ligand interactions. For example, interactive ESTFs between  oxygen atoms in the protein and  nitrogen atoms in the ligand unveil the possible hydrogen bonds, while  interactive ESTFs from protein carbon atoms and ligand carbon atoms indicate hydrophobic effects as shown in Fig. \ref{Figure3}. 
The detailed construction of TFs/ESTFs and the resulting features is described in the following section. 

\subsection{Topological fingerprint and feature construction}
\subsubsection{Correlation functions }

When modeling 3D structure of proteins,  interactions between atoms are related to spatial distances and atomic properties. However, Euclidean metric space does not directly give quantitative description of interaction strengths of  atomic interactions. A nonlinear function is applied to map the Euclidean distances together with atomic properties to a measurement of correlation or interaction between atoms.  Computed atomic pairwise correlation values form a correlation matrix which can then be used to analyze connectivity patterns between clusters of atoms. In the rest of this section, functions that map geometric distance to topological connectivity are referred to as kernels.

\begin{figure}[ht]
\begin{center}
\includegraphics[keepaspectratio,width=0.6\columnwidth]{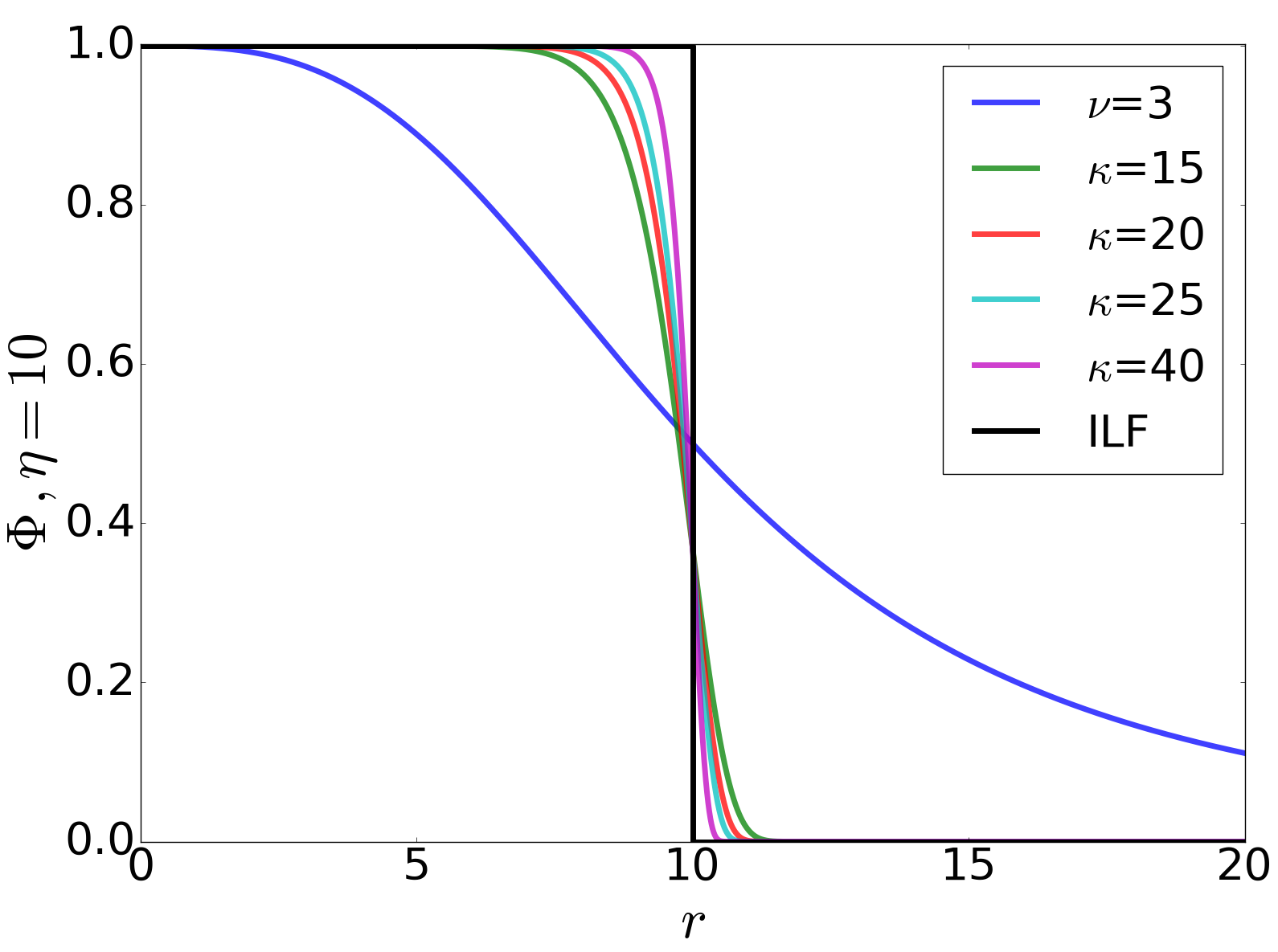}
  \caption{ Illustration of FRI correlation kernels at various powers. At a large  $\kappa$, it essentially becomes an ideal low-pass filter (ILF) and has a cutoff effect. }
  \label{sup:FigureKernel}
\end{center}
\end{figure}

In our previous work, a flexibility-rigidity index (FRI) theory \cite{KLXia:2013d} is introduced which uses decaying radial basis functions to quantify pairwise atomic interactions or correlations. The correlation matrix is then applied to analyze flexibility and rigidity of the protein. The flexibility index computed from the correlation matrix has been found to strongly correlate to experimental B-factors. It has also found its success in the prediction of protein motion \cite{Opron:2014} and the modeling of fullerene stability \cite{KLXia:2015a}. In the previous studies, the most favorable outcomes are obtained with mainly two types of kernels, the exponential kernel and the Lorentz kernel. The exponential kernel is defined as
\begin{equation}\label{eq:exponentialkernel}
\Phi^E(r;\eta_{ij},\kappa) = e^{-(r/\eta_{ij})^\kappa},
\end{equation}
and the Lorentz kernel is defined as 
\begin{equation}\label{eq:Lorentzkernel}
\Phi^L(r;\eta_{ij},\nu) = \frac{1}{1+\left(\frac{r}{\eta_{ij}}\right)^\nu},
\end{equation}
where $k$, $\tau$, and $\nu$ are positive adjustable parameters that control the decay speed of the kernel allowing us to model interactions with difference strengths.  Here $\eta_{ij}$ is the characteristic distance between the $i$th and the $j$th  atoms and is usually set to be the sum of the van der Waals radii of the two atoms.  The correlation between two atoms is then defined as 
\begin{equation}\label{eq:corratom}
C_{ij} = \Phi(r_{ij}),
\end{equation} 
where $r_{ij}$ is the Euclidean distance between the $i$th and the $j$th atoms, and $\Phi$ is the kernel function. Note that the output of the kernel functions lies in the $(0,1]$ interval.  A correlation  matrix is defined as
\begin{equation}\label{eq:metricspace}
d(i,j) = 1-C_{ij}.
\end{equation}
The properties, $\Phi(0,\eta) = 1$, $\Phi(r,\eta)\in (0,1], \forall r\geq 0$, $r_{ij}=r_{ji}$, and the strictly monotone decreasing property of the $\Phi$ assure the  identity of indiscernible, non-negativity, symmetry, and distance increases as pairwise interaction decays. Persistent homology computation is performed with Vietoris-Rips complex built upon the afore-defined correlation  matrix as an addition to the Euclidean space  distance metric.

\subsubsection{TF/ESTF and feature construction}
\begin{table*}[t]
\begin{center}
\caption{Feature extraction from TF/ESTF.}
\begin{tabular}{p{3.8cm} l p{6.5cm}}
\hline
Feature category & TF/ESTF & Features \\ \hline
\multirow{6}{3.8cm}{Connectivity quantified with atomic interaction strengths.} & ${\rm ESTF}(P_{\rm C-C}^{12},{\rm FRI}^{agst}_{\Phi^L/\Phi^E},{\rm VR})$ & \multirow{3}{6.5cm}{Summation of all Betti-0 bar lengths.} \\ 
& $\cdots$ &  \\ 
& ${\rm ESTF}(P_{\rm S-I}^{12},{\rm FRI}^{agst}_{\Phi^L/\Phi^E},{\rm VR})$ &  \\ \cline{2-3}
& ${\rm TF}(P_{\rm all}^6,{\rm FRI}_{\Phi^L/\Phi^E},{\rm VR})$ & \multirow{3}{6.5cm}{Summation of length and birth of Betti-0, -1, and -2 TFs of protein, complex, and difference of the two.} \\
& & \\ 
& ${\rm TF}(P_{\rm pro}^6,{\rm FRI}_{\Phi^L/\Phi^E},{\rm VR})$ & \\ \hline
\multirow{4}{3.8cm}{Physical interactions grouped with intrinsic contact distance.} & ${\rm ESTF}(P_{\rm C-C}^{12},{\rm EUC}^{agst},{\rm VR})$ & \multirow{4}{6.5cm}{Counts of Betti-0 bars with `death' values falling into each interval: $[0,2.5]$, $[2.5,3]$, $[3,3.5]$, $[3.5,4.5]$, $[4.5,6]$, $[6,12]$.} \\ 
 & \multirow{2}{*}{$\cdots$} & \\ 
 & & \\ 
 & ${\rm ESTF}(P_{\rm S-I}^{12},{\rm EUC}^{agst},{\rm VR})$ &  \\ \hline
\multirow{6}{3.8cm}{Geometric features.} & \multirow{3}{*}{${\rm TF}(P_{\rm all}^9,{\rm EUC},{\rm Alpha})$} & \multirow{6}{6.5cm}{Summation of Betti-1 and Betti-2 bar lengths with `birth' value falling into each intervals: $[0,2]$, $[2,3]$, $[3,4]$, $[4,5]$, $[5,6]$, $[6,9]$. The differences between the complex and protein are also taken into account.} \\
 & & \\
 & & \\ 
& \multirow{3}{*}{${\rm TF}(P_{\rm pro}^9,{\rm EUC, Alpha})$} & \\
 & & \\
 & & \\ \hline
\end{tabular}
\label{tb:feature}
\end{center}
\end{table*}

The TFs/ESTFs used in the machine learning process are extracted from the persistent homology computation with a variety of generalized metrics and different groups of atoms. Firstly, the element type and atom center position of the heavy atoms (non-hydrogen atoms) of both protein and ligand molecules are extracted. The hydrogen atoms are neglected because the procedure of completing protein structures by adding missing hydrogen atoms highly depends on the force field chosen which will lead to force field depended effects. The point sets containing certain element types from the protein molecule and certain element types from the ligand molecule are grouped together. With this approach, the interactions between different element types are modeled separately and the parameters that distinguish between the interactions between different pairs of element types can be learned from the training set by  machine learning algorithms. The distance matrices with Euclidean distance and correlation   matrix are constructed for each group of atoms. The features describing the TFs/ESTFs are then extracted from the outputs of persistent homology calculations and glued to form a feature vector for machine learning. The details of TF/ESTF construction and feature generation follow in the rest of this section.

\paragraph{Groups of atoms.} The  types of elements considered  for proteins are ${\rm T}_{\rm P}=\lbrace {\rm C, N, O, S} \rbrace$ and those for ligands are ${\rm T}_{\rm L}=\lbrace {\rm C, N, O, S, P, F, Cl, Br, I}\rbrace$. 
We denote $P_{\rm X-Y}^c$ a set of atoms that consist of ${\rm X}$ type of atoms in protein and ${\rm Y}$ type of atoms in ligand, and the distance between any pair of atoms in these two groups is within a cutoff $c$: 
\begin{equation}\label{eq:atomgroup}
P_{\rm X-Y}^c = \lbrace a |a\in {\rm X},  \min\limits_{b\in {\rm Y}}{\rm dis}(a,b)\leq c \rbrace\cup\lbrace b|b\in {\rm Y} \rbrace,
\end{equation} 
where $a$ and $b$ denote atoms.  As an example, $P_{\rm C-O}^{12}$ contains all ${\rm O}$ atoms in the ligand and all ${\rm C}$ atoms in the protein  that are within the cutoff distance of $12$ \AA~from the ligand molecule. We denote the set of all heavy atoms in ligand together with all heavy atoms in protein that are within the cutoff distance $c$ from the ligand molecule by $P_{\rm all}^c$. Similarly, the set of all  heavy atoms  in protein that are within the cutoff distance $c$ from the ligand molecule by  $P_{\rm pro}^c$.

\paragraph{Distance matrices.} We define FRI based correlation matrix and Euclidean metric based distance matrix in this section. The Lorentz kernel and exponential kernel defined in Eq.\ref{eq:Lorentzkernel} and Eq.\ref{eq:exponentialkernel}, see Fig. S\ref{sup:FigureKernel}, are used due to their success in our previous studies of biomolecules with FRI theory. We use $A(i)$ to denote the affiliation of the atom with index $i$ which is either  protein or ligand.
\begin{itemize}
\item ${\rm FRI}_{\tau,\nu}^{agst}$:
\begin{equation}\label{eq:FRI_against}
d(i,j) = 
\begin{cases}
1 - (\frac{1}{1+( r_{ij}/\eta_{ij})^\nu}), \: &A(i) \neq A(j) \\
d_{\infty}, &A(i)=A(j)
\end{cases}
\end{equation}
where $r_{ij}$ is the Euclidean distance between atoms with indices $i$ and $j$ and $\eta_{ij}=\tau*(r_i+r_j)$. The superscript $agst$ is the abbreviation of against and means that only the interaction between atoms in the protein and atoms in the ligand are taken into account since the binding between protein and ligand is studied and thus the distance between atoms from the same molecule are set to $d_{\infty}$ which is a large positive number.

\item ${\rm FRI}_{\tau,\nu}$:
\begin{equation}\label{eq:FRI}
d(i,j) = 1 - \left(\frac{1}{1+( r_{ij}/\eta_{ij})^\nu}\right)
\end{equation}

\item ${\rm EUC}^{agst}$:
\begin{equation}\label{eq:EUC_against}
d(i,j) = 
\begin{cases}
r_{ij}, \: & A(i)\neq A(j) \\
d_{\infty}, & A(i)=A(j)
\end{cases}
\end{equation}
\item ${\rm EUC}$:
\begin{equation}\label{eq:EUC}
d(i,j) = r_{ij}.
\end{equation}
\end{itemize}

The distance matrices with distance functions ${\rm FRI}^{agst}_{\tau,\nu}$ and ${\rm EUC}^{agst}$ mainly model connectivity and interaction between protein and ligand molecules and even higher order correlations between atoms. The Euclidean metric space is applied to detect geometric characteristics such as cavities and cycles. 

We denote the output of persistent homology computation by TF$(x,y,z)$ where $x$ is the set of atoms, $y$ is the distance matrix used, and $z$ is the simplicial complex constructed. The notation ESTF$(x,y,z)$ is used if $x$ is element specific. The extraction of machine learning feature vectors from TFs is summarized in Table \ref{tb:feature}.

The division of Betti-0 bars into bins is because that different categories of atomic interactions have their distinguish intrinsic distances such as 2.5 \AA~for ionic interactions and 3 \AA~for hydrogen bonds. The separation of Betti-1 and Betti-2 TFs helps grouping the geometric features of various scales. For FRI kernel, different pairs of parameters $\tau$ and $\nu$ are used to characterize interaction of different scales. In this work, the specific pairs of $[\tau,\nu]$ used are $[0.5,3]$, $[1,3]$, and $[2,3]$.

\section{Results}
To demonstrate the power of the proposed topological learning strategies for protein-ligand binding affinity prediction, we consider  benchmark test sets from the PDBBind database \cite{PDBBind:2015}. This database provides a comprehensive collection of binding affinity data of biomolecular complexes available in \href{www.rcsb.org}{Protein Data Bank (PDB)} \cite{Berman:2000} that are obtained from experiments. 
 We first consider  the PDBBind v2007 refined set of 1300 protein-ligand complexes, which include the v2007 core set of 195 complexes \cite{RenxiaoWang:2009Compare}. Each protein-ligand complex  has the 3D structure of the biomolecule accompanied by the 3D structure of the ligand and the  experimental binding affinity. The structures are directly imported to our topological learning predictor without any structure optimization, which tests the robustness of the predictor over data of relatively low quality. The v2007 core set has been used to evaluate the performance of more than 20 exiting scoring functions \cite{Li:2015,  IDScore:2013,   RenxiaoWang:2009Compare} 
We use the refined set, excluding the core set, of 1105 complexes as the training set, while the core set of 195 complexes is used as the test set in our study.


\begin{figure}[ht]
\begin{center}
\includegraphics[keepaspectratio,width=0.7\columnwidth]{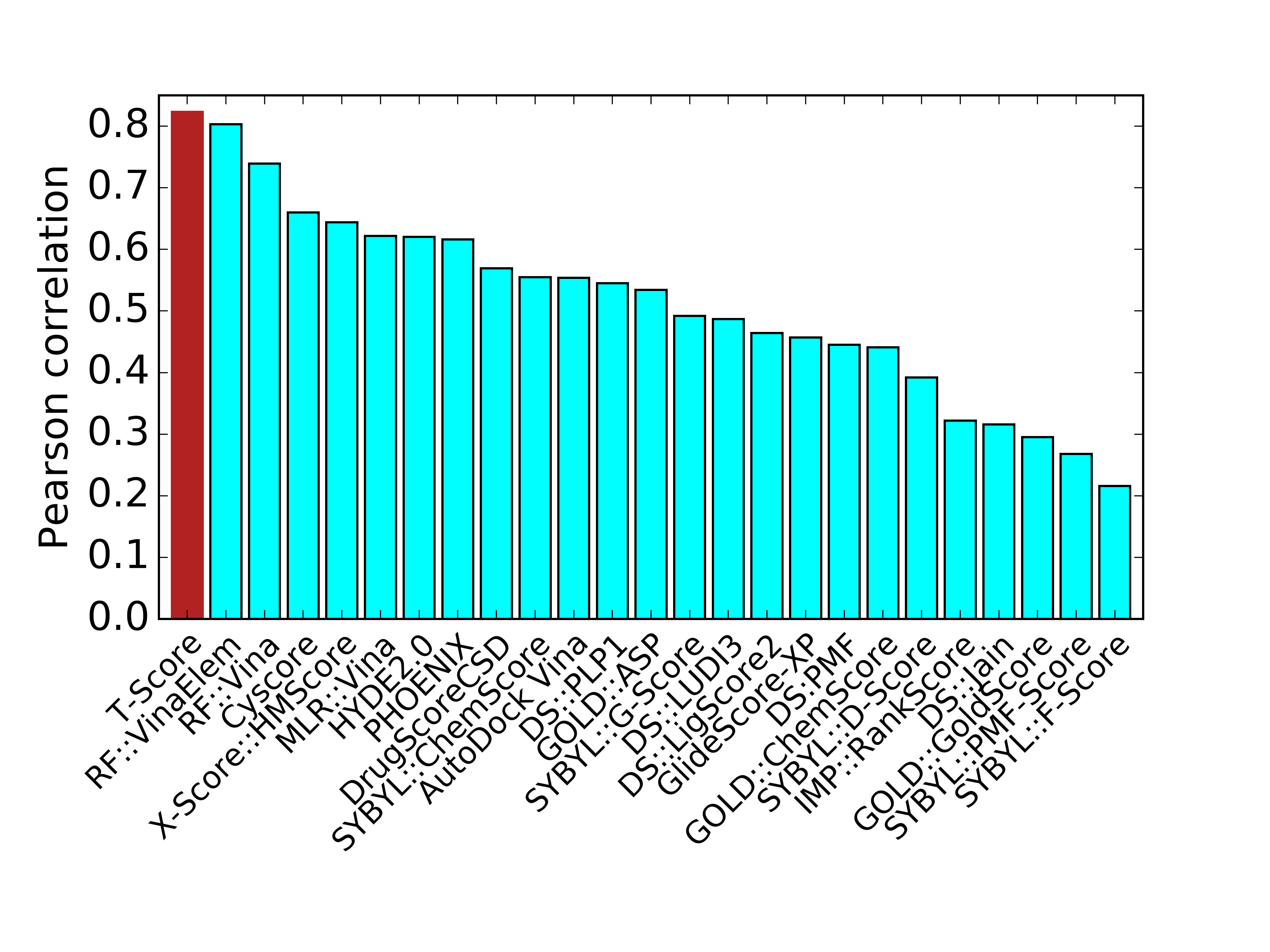}
\caption{Comparison of Pearson correlation coefficients of various predictions and experimental binding affinity data for the PDBBind 2007 core set of 195 complexes. Topological score (T-Score) constructed from the combination of  machine learning and persistent topology outperforms all the other eminent scoring functions. Its  Pearson correlation coefficient is 0.824 with an RMSE of 1.881 kcal/mol.}
\label{Figure5}
\end{center}
\end{figure}

Figure  \ref{Figure5} illustrates a comparison of Pearson correlation coefficients of  the present topological-score (T-Score) and other 24 scoring functions given in the literature \cite{Li:2015}. Amazingly,  the proposed topological learning strategy outperforms all other methods, which represented the state of the art. The strong performance of the proposed topological learning strategy indicates the  power of TFs/ESTFs for revealing protein-ligand binding mechanism, which is further explored in the rest of this paper. 

\begin{table}[ht]
\caption{The overall performance of the T-score binding affinity predictor on a variety of testing tasks. The average performance is reported with the best performance in the parenthesis. PCC is the Pearson correlation coefficient and RMSE is the root mean squared error in the unit of Kcal/mol. CV stands for cross-validation}
\begin{center}
\begin{tabular}{cccc}
\hline
Dataset & Task & PCC & RMSE \\ \hline
v2007 refined & 5-fold CV & 0.750 (0.765) & 1.945 (1.894) \\ \hline
v2007 refined & v2007 core & 0.817 (0.824) & 1.919 (1.881) \\ \hline
v2013 refined & 5-fold CV & 0.750 (0.758) & 1.801 (1.779) \\ \hline
v2013 refined & v2013 core & 0.766 (0.772) & 2.067 (2.053) \\ \hline
\end{tabular}
\label{tb:featureperformance2}
\end{center}
\end{table}
      
It remains to verify that the outstanding performance of ESPH based binding affinity predictions is not limited to a specific data set. To this end, we further consider the PDBBind v2013 core set of 195 protein-ligand complexes. Our training was carried out over the PDBBind v2013 refined set of 2959 entries, excluding the v2013 core set \cite{PDBBind:2015}. This refined set was  selected from a general set of 14,620 protein-ligand complexes with good quality, filtered over  binding data, crystal structures, as well as the nature of the complexes  \cite{PDBBind:2015}. The PDBBind v2013 core set has been tested by  a large number of computational methods  \cite{HongJianLi:2015}. The results for the tasks of predefined training testing split and 5-fold cross validation on PDBBind v2013 refined set and PDBBind v2007 refined set are summarized in Table \ref{tb:featureperformance2}.
\section{Discussion}
A crucial attribute to the present cutting edge binding prediction is the characterization of interactions. It is well-known that one can categorize hydrogen bonds with donor-acceptor distances of 2.2-2.5\AA~ as strong, mostly covalent, 2.5-3.2\AA~ as moderate, mostly electrostatic, and 3.2-4.0\AA~  as weak, electrostatic  \cite{Jeffrey}. Their corresponding energies are about 40-14, 15-4, and less than 4 kcal/mol, respectively \cite{Jeffrey}. Motivated by this observation, we introduce binned barcode representation (BBR) to distinguish  different bond  distances  and thus, corresponding energies. Specifically, we divide Betti-0 ESTFs into various bins,  namely [0, 2.5], (2.5, 3], (3, 3.5], (3.5, 4.5], (4.5, 6] and (6, 12]\AA~ for the generation of features. This approach enables us to precisely characterize  hydrogen bonds, and hydrophilic and hydrophobic interactions. 

It is interesting to analyze the predictive powers of ESTFs. Typically, a more important ESTF has a higher  predictive power. We first examine the performance of Betti-0 ESTFs of non-carbon heavy atoms that correspond to hydrogen bonds or hydrophilic interactions. This set of Betti-0 ESTFs delivers a Pearson correlation coefficient of 0.63 with a root-mean-square error (RMSE) of 2.51 kcal/mol, which is better than results  of most other methods shown in Fig.  \ref{Figure5}.  
In contrast, Betti-0 ESTFs of carbon atoms only achieve a Pearson correlation coefficient of 0.59 with RMSE of 2.62 kcal/mol. However, the mechanism of hydrophobic interactions differs much from that of hydrogen bonds, which are electrostatic in origin. Protein hydrophobic interactions are strengthened by the aggregation of nonpolar carbon atoms. Such aggregation is effectively detected by carbon Betti-1 and Betti-2 ESTFs. Indeed,  Betti-1 and Betti-2 ESTFs alone give rise to a Pearson correlation coefficient of 0.76 with an RMSE of 2.14 kcal/mol. The combination of all carbon ESTFs yields  a respectable Pearson correlation coefficient of 0.79 with an RMSE of 2.04 kcal/mol. Therefore, hydrophobic interactions represented by carbon ESTFs have a higher predictive power than hydrophilic interactions represented by the ESTFs of non-carbon heavy atoms. The combination of Betti-0 ESTFs of all heavy-atoms renders a remarkable Pearson correlation coefficient of 0.81 with RMSE of 1.97 kcal/mol. This result indicates that  hydrophilic and hydrophobic interactions are complementary but not additive to each other.  

The aforementioned results are obtained from the ESTFs based on the standard Euclidean space distance filtration. Alternatively, a correlation function based filtration can be utilized \cite{KLXia:2014c}. Specifically, the correlation function measures the distance between each pair of atoms with a specific scale so that certain distance is favored. One of correlation functions successfully used in protein flexibility analysis is the Lorentz function $\left[1+\left(\frac{r_{ij}}{\eta_{ij}} \right)^{\nu}\right]^{-1}$, where  $r_{ij}$ is the  distance between atom $i$ and atom $j$,  and $\eta_{ij}=\tau(r_i+r_j)$ is the scale \cite{KLXia:2013d}. Here $r_i$ and $r_j$ are the van der Waals radii of atom $i$ and atom $j$, respectively.  Obviously, this filtration is regulated by a pair of parameters, i.e., the correlation power $\nu$ and the correlation scale $\tau$. It is found that Betti-0 ESTFs constructed from Lorentz function based filtration are able to accurately predict binding affinities. Specifically, when $(\nu,\tau)=(3,0.5),(3,1),(3,2), (2,1)$, and $(5,1)$, the prediction Pearson correlation coefficients are 0.762, 0.758, 0.769, 0.777 and 0.772, respectively. Obviously, other forms of correlation function can be used as well. For example, Betti-0 ESTFs constructed from the exponential function  ${\rm exp}\left[ - \left(\frac{r_{ij}}{\eta_{ij}}\right)^\kappa\right]$ based filtration with $\kappa=\tau=1$ delivers a  Pearson correlation coefficient of 0.782. An interesting advantage of correlation function based filtration is that multiscale effects in protein-ligand interactions can be captured by the use of multiple sets of ESTFs characterized  at different scales.  The combination of three sets of Betti-0 ESTFs  $(\nu,\tau)=$(3,0.5),(3,1) and (3,2) leads to a high   Pearson correlation coefficient of 0.784. 

 It is noted that the  T-Score result depicted in Fig. \ref{Figure5} is the best value obtained from a combination of binned Betti-0 TFs of all heavy atoms, Betti-1 and Betti-2 TFs of all heavy atoms which are able to capture binding pockets and cavities,  Betti-1 and Betti-2 ESTFs of carbon atoms and finally, three sets of element specific Betti-0 ESTFs generated from  Lorentz correlation functions at  $(\nu,\tau)=(3,0.5),(3,1)$ and $(3,2)$. The median   Pearson correlation coefficient from 500 random tests is as high as 0.817 with an RMSE of 1.912 kcal/mol.

\begin{figure}[ht]
\begin{center}
\includegraphics[keepaspectratio,width=0.6\columnwidth]{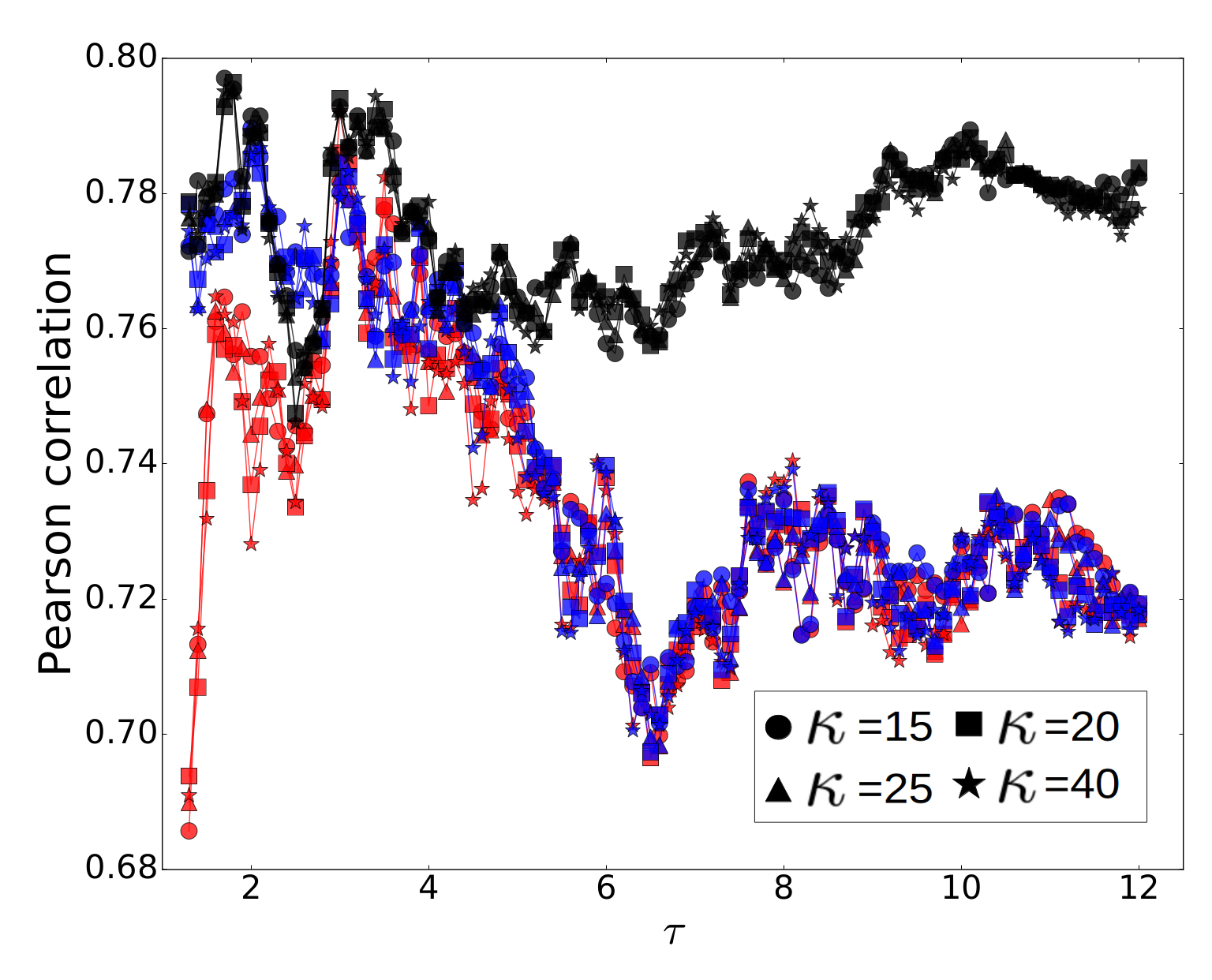}
\caption{The Pearson correlation coefficient of Type I (red), Type II (blue) and Type III (black) features with various cutoff distances achieved by varying the length scale $\tau$   in predicting the PDBBind v2007 core set. 
All Betti-0 features are derived from VR complexes with the exponential correlation function with $\kappa=15,20,25$ or $40$. The Betti-1 and Betti-2 features are extracted from  alpha complexes of protein-ligand ${\rm C-C}$ networks.  The training/testing process is performed 50 times for each data point and the average is taken.
}
\label{fig:DistanceEffect}
\end{center}
\end{figure}

Finally, we demonstrate that the present topological learning strategy is able to bring to light the effective lengths of various interactions.  To this end, we consider the exponential function  based filtration with a  large $\kappa$ value, i.e., $\kappa=15,20,25$ or $40$,  which results in an effective cutoff at the length scale of $\eta_{ij}=\tau (r_i+r_j)$ \cite{KLXia:2015f}. Meanwhile, we consider three types of features, as shown in Fig. \ref{fig:DistanceEffect}.   Type I shown in red includes  all the Betti-0 features except those from protein-ligand ${\rm C-C}$ pairs. Type II shown in blue is obtained by restoring in Type I the Betti-0 features from protein-ligand ${\rm C-C}$ pairs. Type III shown in black is created by adding protein-ligand ${\rm C-C}$ Betti-1 and Betti-2 features to Type II.  Interestingly,  Fig. \ref{fig:DistanceEffect} reveals molecular  mechanism about protein-ligand binding. First, there are oscillatory peaks in Pearson correlation coefficients around $\tau=1.6,$ and $3.1$, indicating the importance of the complete inclusion of the first layer and/or the second layer of residues in predictive models. Additionally, it is interesting to note the importance of protein-ligand ${\rm C-C}$ interactions  in nearest two layers of residues to binding. However, in the range of 3-5$\tau$, the inclusion of protein-ligand ${\rm C-C}$ interactions appears to have little effect on binding predictions. Finally, at large length scales ($\tau>5$), protein-ligand ${\rm C-C}$ Betti-1 and Betti-2 features give rise to a surprising catch-up contribution to binding predictions. This finding indicates that  protein-ligand ${\rm C-C}$ network or protein-ligand hydrophobic effect is still significantly effective at more than 40\AA~ away from the binding site, which has an important ramification to drug design and protein design.

 \section*{Acknowledgement}
 This work was supported in part by NSF Grants DMS-1160352 and IIS-1302285     and
MSU Center for Mathematical Molecular Biosciences Initiative.  

%

\bibliographystyle{ieeetr}
\bibliography{refs}

\end{document}